\documentclass[reprint,aps,prl,showpacs,superscriptaddress]{revtex4-1}
\usepackage{amssymb,amsmath}
\usepackage{graphicx}
\usepackage{dcolumn}
\usepackage{multirow}
\usepackage{color}
\usepackage[english]{babel}
\usepackage{hyperref}

\newcommand{\ket}[1]{ |{#1} \rangle }

\begin{document}

\def\simlt{\mathrel{\lower .3ex \rlap{$\sim$}\raise .5ex \hbox{$<$}}}
\def\simgt{\mathrel{\lower .3ex \rlap{$\sim$}\raise .5ex \hbox{$>$}}}

\title{\textbf{\fontfamily{phv}\selectfont Pulse-gated quantum dot hybrid qubit}}
\author{Teck Seng Koh}
\author{John King Gamble}
\author{Mark Friesen}
\author{M. A. Eriksson}
\author{S. N. Coppersmith}
\affiliation{Department of Physics, University of Wisconsin-Madison, Madison, WI 53706}



\begin{abstract}
A quantum dot hybrid
qubit formed from three electrons in a double quantum dot has the potential for great speed, due to presence of level crossings where the qubit becomes charge-like.
Here, we show how to take full advantage of the level crossings in a pulsed gating scheme, which decomposes the spin qubit into a series of charge transitions.
We develop one and two-qubit dc quantum gates that are simpler than the previously proposed ac gates.
We obtain closed form solutions for the control sequences and show that these sub-nanosecond gates can achieve high fidelities.
\end{abstract}

\maketitle

\begin{figure}[t]
\includegraphics[width=2.8in]{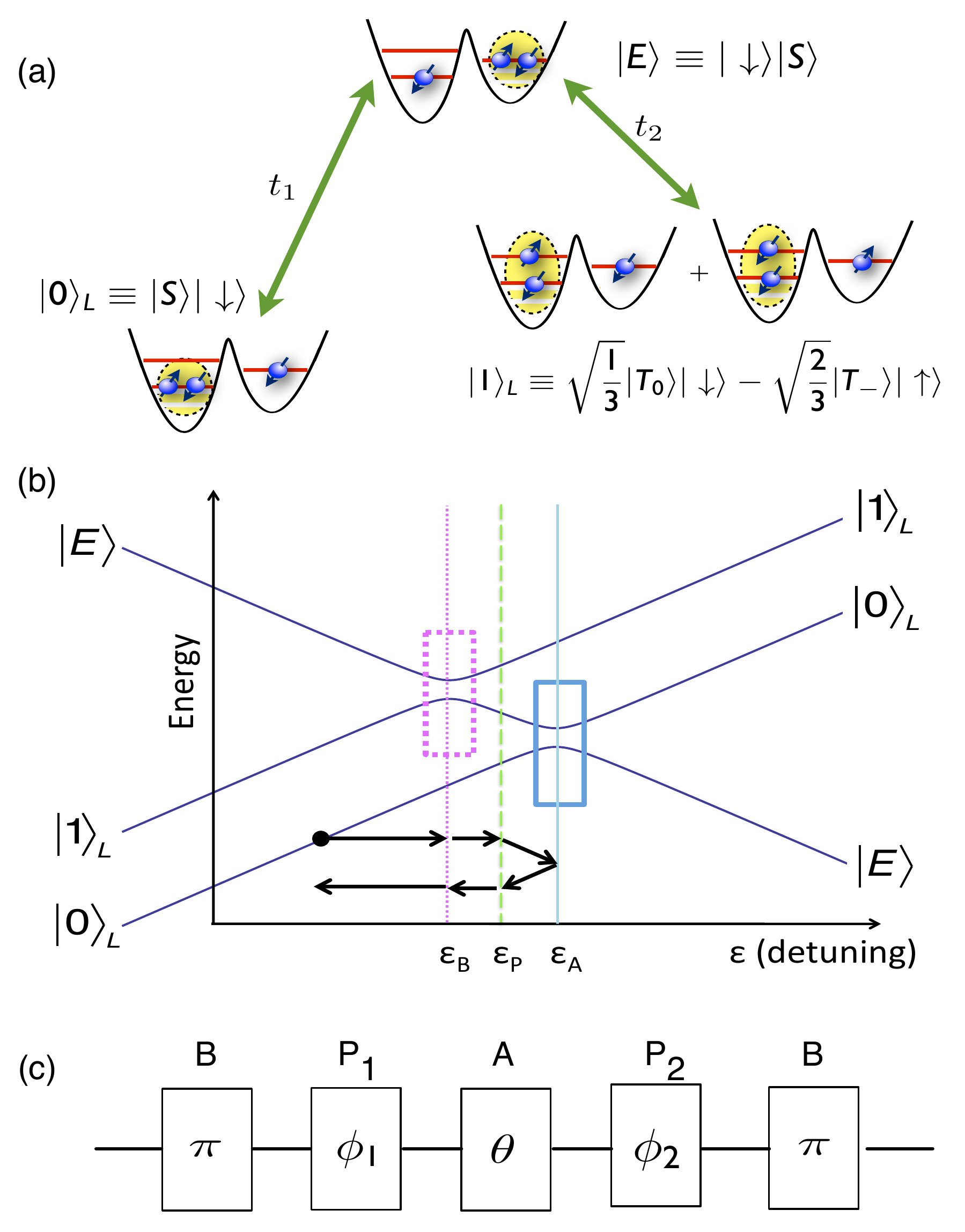}
\caption{\label{fig:combined} 
(a) Schematic of the quantum dot hybrid qubit and of the physics
underlying gate operations.
The logical qubit states are
$\ket{0}_L=\ket{S}\ket{\downarrow}$ and
$\ket{1}_L=\sqrt{\frac{1}{3}}\ket{T_0} \ket{\downarrow}
- \sqrt{\frac{2}{3}}\ket{T_{-}}\ket{\uparrow}$, where
$\ket{S}$,
$\ket{T_-}$, and 
$\ket{T_0} $
are two-particle singlet (S) and triplet (T) states in the left dot, and
 $\ket{\uparrow}$ and $\ket{\downarrow}$ respectively
denote a spin-up  and spin-down
 electron in the right dot.
Introducing tunneling with amplitudes $t_1$ and
$t_2$ to an intermediate excited state $\ket{E}$
with one electron in the left dot and
two electrons in the right dot induces transitions
between $\ket{0}_L$ and
$\ket{1}_L$.
(b)  Energies of the three relevant states $\ket{0}_L$, $\ket{1}_L$, and $\ket{E}$
as a function of the detuning $\varepsilon$ between the two dots.
The ground state has  two
electrons in the left dot when $\varepsilon<\varepsilon_A$ and
two electrons in the right dot when $\varepsilon>\varepsilon_A$; 
the qubit operates mainly in the regime $\varepsilon \le \varepsilon_A$.
The energy difference between the qubit states
 $\ket{0}_L$ and $\ket{1}_L$ is large for all values of $\varepsilon$, but
there is an avoided crossing between $\ket{0}_L$ and $\ket{E}$
at the detuning value $\varepsilon_A$ (blue box),
and another avoided crossing between 
$\ket{1}_L$ and $\ket{E}$
at the detuning value $\varepsilon_B$ (dotted magenta box).
Pulse-gate transitions between $\ket{0}_L$ and $\ket{1}_L$ can
be performed by using both avoided crossings. 
Pulses to the detuning value $\varepsilon_P$ are used in the gating
scheme to induce phase differences between the three states.
A gating sequence to provide arbitrary rotations between the logical qubits states $\ket{0}$ and $\ket{1}$ is  indicated with arrows at the bottom;
along the detuning axis, the pulse sequence is given by 
$\varepsilon_\text{init}\rightarrow\varepsilon_B\rightarrow\varepsilon_P\rightarrow\varepsilon_A\rightarrow\varepsilon_P\rightarrow\varepsilon_B\rightarrow\varepsilon_\text{final}$.
(c) The corresponding circuit diagram of the gate sequence, with time progressing from left to right.
Gates $P_1$, $A$, and $P_2$ are tunable, with the control parameters $\phi_1$, $\theta$, and $\phi_2$ given in Eqs.~\ref{eq:theta}-\ref{eq:phi2}.
}
\end{figure}

A key figure of merit for a quantum information processing device is the ratio
of the quantum coherence time to the time required
to perform qubit gate manipulations~\cite{DiVincenzo:1995p2698,Preskill:1998p469,Fisher:2003p1441}.
The recently proposed hybrid quantum dot qubit~\cite{Shi:2012p140503} is a relatively simple qubit
architecture that could achieve
a higher figure of merit than previous qubit designs~\cite{Koppens:2006p766,Levy:2002p1446,DiVincenzo:2000p1642}.
The qubit itself is a set of two states with total spin quantum numbers
$S^2 = 3/4~(S=\frac{1}{2})$ and $S_z = - \frac{1}{2}$, with the two different states using the singlet and triplet in a doubly-occupied dot and a single spin in a singly-occupied dot.
The two states of the qubit have different energies, and
Ref.~\onlinecite{Shi:2012p140503} proposes to implement gate operations
using high-frequency ($\sim$10$-$40$~{\rm GHz}$) resonant RF pulses.
This method is feasible~\cite{Martinis:2002p117901,PioroLadriere:2008p776}, but it is significantly
more complicated to implement experimentally than the pulse-gating methods
used for charge qubits in Refs.~\onlinecite{Hayashi:2003p226804,Gorman:2005p1397,Petta:2006p474,Shinkai:2009p056802,Petersson:2010p246804}
and for spin qubits
in Refs.~\onlinecite{Petta:2005p2180,Laird:2010p1985,Gaudreau:2011p54,Shulman:2012p202,Maune:2012p344}.
Here we show how to implement pulse-gating of the quantum dot
hybrid qubit.  One- and two-qubit gates require a modest number of
 non-adiabatic voltage pulses (five and eight, respectively), 
 each of which is similar to those already used for gate operations on charge qubits and singlet-triplet spin quits.

The two logical qubit states of the hybrid quantum dot qubit 
are $\ket{0}_L=\ket{S}\ket{\downarrow}$ and
$\ket{1}_L=\sqrt{\frac{1}{3}}\ket{T_0} \ket{\downarrow}
- \sqrt{\frac{2}{3}}\ket{T_{-}}\ket{\uparrow}$, where
$\ket{S}$,
$\ket{T_-}$, and 
$\ket{T_0} $
are two-particle singlet (S) and triplet (T) states in the left dot, and
 $\ket{\uparrow}$ and $\ket{\downarrow}$ respectively
denote a spin-up  and spin-down
 electron in the right dot.
These states form
a decoherence-free subspace
that
is insensitive to long-wavelength magnetic flux noise; moreover,
decoherence processes that do not explicitly couple to spin 
or induce a transition of an electron to the reservoir
do not induce transitions that go outside of the subspace of an individual qubit~\cite{Lidar:1998p2594}.
The qubit has the same symmetries in spin space as the triple-dot
qubit proposed by DiVincenzo et al.~\cite{DiVincenzo:2000p1642}, but
is simpler to fabricate because it requires a double dot instead of a triple dot.
Transitions between the logical qubit states $\ket{0}_L$ and $\ket{1}_L$
are allowed when tunneling is introduced
between the dots.
The physical process that
leads to transitions between the two logical qubit states $\ket{0}_L$
and $\ket{1}_L$ involves an intermediate state $\ket{E}$ that
has one electron in the left dot and 
two electrons in the right dot, and the same
total $S^2$ and $S_z$.
Fig.~\ref{fig:combined}(a) is a schematic of the hybrid qubit and of
a physical process that
yields transitions between the logical states $\ket{0}_L$ and $\ket{1}_L$.
In the figure the doubly occupied right dot is labeled as having a singlet
ground state,
but the energy level diagram applies for both positive
and negative singlet-triplet energy splittings in the right dot, and the spin of the lower energy state is not essential for the discussion below.
We assume that the singlet-triplet splitting in the right dot is large enough
that higher energy states of the right dot do not mix appreciably with the states
that we consider explicitly here.

Rabi oscillations between two quantum states $\ket{\alpha}$
and $\ket{\beta}$ are achieved by changing the detuning
suddenly to a value at which
the energy difference between the states is smaller than the coupling between them.
Very near the avoided crossing between two states,
the time evolution is prescribed by the two-state Hamiltonian
\begin{equation}
H=\left (
\begin{array} {cc}
\tilde{\varepsilon} & \Delta \\
\Delta & -\tilde{\varepsilon}
\end{array}
\right )~,
\end{equation}
where $\Delta$ is the coupling between the two
states $\ket{\alpha}$
and $\ket{\beta}$
and $2\tilde{\varepsilon}$ is the energy difference between the two states in the
absence of coupling.
Significant mixing between the states occurs only when $\tilde{\varepsilon} \simlt \Delta$.
If one pulses the system suddenly to $\tilde{\varepsilon}=0$, so that the state at time $t=0$ is
$\ket{\psi(0)}=\ket{\alpha}$, then the time evolution of the two-state system is given by
$\ket{\psi(t)} = \cos(\Omega_R t) \ket{\alpha} - i \sin(\Omega_R t) \ket{\beta}$, which oscillates between $\ket{\alpha}$ and $\ket{\beta}$ at the Rabi frequency 
$\Omega_R=\Delta/\hbar$.
A pulse of duration $T$ rotates the state on the Bloch sphere by an angle
$\Theta = 2\Omega_R T$ around the $x$-axis~\cite{Nielsen:2000}.

Fig.~\ref{fig:combined}(b) shows the energies of the states
$\ket{0}_L$, $\ket{1}_L$, and $\ket{E}$ as
a function of detuning.
The energy difference between $\ket{0}_L$ and $\ket{1}_L$, which is
 the singlet-triplet energy splitting in the left dot, typically is substantial (of order $0.1$~meV,
corresponding to a frequency $\sim$25 GHz)
and depends only
moderately on the detuning~\cite{Shi:2011p233108}, so achieving an avoided
crossing of $\ket{0}_L$ and $\ket{1}_L$ is typically not feasible.
Therefore,
pulse-gating is
ineffective in inducing transitions
directly between the two qubit states.
However, there is a value of the detuning $\varepsilon_A$ at which there is an avoided
crossing between the states $\ket{E}$ 
and $\ket{0}_L$, and another value of the detuning
$\varepsilon_B$
at which there is an avoided crossing between the
states $\ket{E}$ and $\ket{1}_L$.
Transitions from $\ket{0}_L$ to $\ket{1}_L$ can be induced
by first pulsing to $\varepsilon_A$, the avoided crossing between
$\ket{0}_L$and $\ket{E}$, and then pulsing to
 $\varepsilon_B$, the avoided crossing between $\ket{E}$ and $\ket{1}_L$.
Similarly, transitions from $\ket{1}_L$ to $\ket{0}_L$ can be induced by
first pulsing to $\varepsilon_B$ 
and then pulsing to $\varepsilon_A$.
These arguments show how to induce transitions from $\ket{0}_L$
 to $\ket{1}_L$ and from $\ket{1}_L$ to $\ket{0}_L$.
 
 \emph{Arbitrary Rotations of One Hybrid Qubit.}---We now present a pulse sequence that implements $U(\hat{\mathbf{n}},\beta)$, a rotation of the logical qubit by an angle
$\beta$ about the rotation axis 
 $\hat{\mathbf{n}}=(\sin \eta \cos \zeta , \sin \eta \sin \zeta,\cos \eta)$,
where $\eta$ and $\zeta$ are the polar and azimuthal angles.
The sequence is constructed from three primitive gates: A, B, and P.
The $B$ gate is implemented by pulsing the detuning parameter to $\varepsilon_B$ for
a time that results in
a $\pi$ rotation about the $x$ axis in the $\{ \ket{1}_L,\ket{E} \}$ subspace, thus converting
$\ket{1}_L \rightarrow \ket{E}$ and $\ket{E} \rightarrow \ket{1}_L$.
The $A$ gate, obtained by pulsing to $\varepsilon_A$ for an amount of time that
implements a rotation 
by an arbitrary rotation angle $\theta$
about the $x$ axis in the $\{ \ket{0}_L,\ket{E} \}$ subspace,
changes the ``latitude" of the qubit on the $\{ \ket{0}_L,\ket{E} \}$ Bloch sphere.
The ``longitude" on the $\{ \ket{0}_L,\ket{E} \}$ Bloch sphere is controlled using
a phase gate $P$ that is obtained by pulsing to a detuning $\varepsilon_P$
between the anticrossings at $\epsilon_A$ and $\epsilon_B$,
as shown in Fig.~\ref{fig:combined}(b), at which
state $\ket{E}$ gains a phase $\phi$ relative to $\ket{0}_L$.
The phase gate $P$ is very fast, due to the large energy difference between $\ket{0}_L$ and $\ket{E}$.
By inserting two phase gates that rotate the phase
by angles $\phi_1$ and $\phi_2$, between the $B$ and $A$ gates, any prescribed rotation on the $\{ \ket{0}_L,\ket{1}_L \}$ Bloch
sphere can be obtained.
The full pulse sequence is shown on the detuning axis at the bottom of Fig.~\ref{fig:combined}(b) and is
also shown as a circuit diagram in Fig.~\ref{fig:combined}(c), corresponding to the gate sequence $U=BP_2AP_1B$.
The relationship between the rotation parameters $(\beta,\eta,\zeta)$ and the control parameters $(\theta,\phi_1,\phi_2)$, derived in the Supplemental Information, is:
\begin{gather}
\theta = 2\arcsin [\sin (\eta) \sin (\beta/2) ] , \label{eq:theta} \\
\phi_1 =  \arctan [\cos (\eta) \tan (\beta/2) ]-\phi_B -\zeta +\pi/2 ,\\
\phi_2 =  \arctan [\cos (\eta) \tan (\beta/2) ]-\phi_B +\zeta +\pi/2 ,  \label{eq:phi2}
\end{gather}
where $\phi_B$ is the incidental phase gained by state $\ket{1}$ relative to $\ket{0}$ while implementing the $B$ gate.
For example, an $x$ rotation with angle $\beta$ is obtained from the sequence
$U=BP_2(-\phi_B+\pi/2)A(\beta)P_1(-\phi_B+\pi/2)B$.
In this case, we can view the action of the $P$ gates as simply removing the phase gained during the $B$ gates.
 
The speed of a pulsed gate in a quantum dot hybrid qubit can
be estimated by noting that it is composed
of five primitive gates, as shown in Fig.~\ref{fig:combined}(c).  
The $A$ and $B$ gates correspond to charge qubit rotations, and their speed is determined by the anti-crossing energy gaps~\cite{Hayashi:2003p226804,Gorman:2005p1397,Petta:2004p1586,Shinkai:2009p056802,Petersson:2010p246804}.
A $\pi/2$ rotation of a charge qubit can be implemented in a time
$\simlt$200 ps~\cite{Petersson:2010p246804}.
Gates $P_1$ and $P_2$ are phase gates, and their speed is determined by the energy splitting between states $\ket{0}$ and $\ket{E}$.  
For a splitting of 50~$\mu$eV, a single $P$ gate can be implemented in $\leq 80$~ps.
Thus, sub-nanosecond gating of a hybrid-qubit should be achievable with current
technology.

\begin{figure}[t]
\includegraphics[width=2.8in]{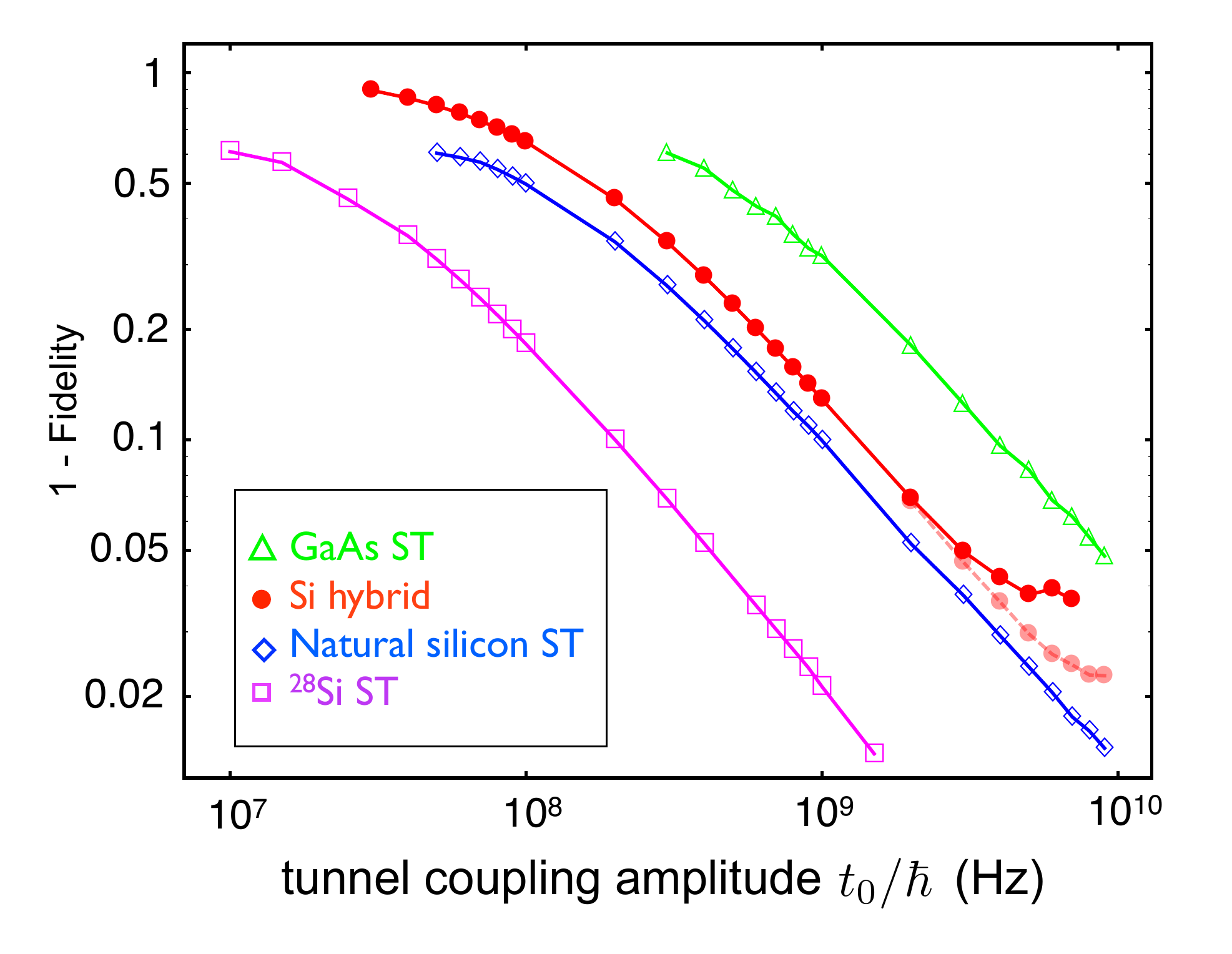}
\vspace{-8pt}
\caption{\label{fig:fidelity} 
Numerical calculations of the infidelity (1-fidelity) caused by dephasing during qubit rotations
without use of dynamical decoupling~\cite{Viola:1999p2417}, 
as a function of the inter-dot tunnel coupling in a double quantum dot.
Solid red circles show the results for a hybrid qubit, and
open triangles, diamonds, and squares show the results for the exchange gate of
singlet-triplet qubits~\cite{Petta:2005p2180,Maune:2012p344} for several different physical systems:  GaAs, natural Si and isotopically purified Si.
The two curves for the Si hybrid qubit represent different values of the singlet-triplet
splitting (0.05 meV for the upper curve and 0.5 meV for the lower curve).
The fidelity of the pulse-gated hybrid qubit is comparable to that of a singlet-triplet exchange gate in natural
silicon.
}
\end{figure}

\emph{Decoherence.}---When hybrid qubits are not undergoing gate operations, their coherence properties benefit from their spin-like character, similar to singlet and triplet states in a two-electron quantum dot~\cite{Gamble:2012p035302}.
However, the gating procedures described above consist of sequential rotations of charge qubits, for which the decoherence rates are faster.
The gating speeds are also faster, so realistic estimates for the gate fidelity require us to perform dynamical simulations of the gate sequence.

We model the dynamical evolution of the density matrix $\rho$ using a 
master equation~\cite{Nielsen:2000}:
$i\hbar \dot{\rho}=[H,\rho]+D$.
The Hamiltonian and decoherence terms are expressed in the $\{ \ket{0}_L,\ket{1}_L,\ket{E} \}$ basis as
\begin{equation*}
H=\begin{pmatrix}
0 & 0 & t_1 \\
0 & E_{01} & -t_2 \\
t_1 & -t_2 & -\varepsilon
\end{pmatrix} \hspace{.05in} \text{and} \hspace{.05in} 
\frac{iD}{\hbar}=\begin{pmatrix}
0 & \gamma \rho_{01} & \Gamma \rho_{0E} \\
\gamma \rho_{10} & 0 & \Gamma \rho_{1E} \\
\Gamma \rho_{E0} & \Gamma \rho_{E1} & 0 \end{pmatrix} .
\end{equation*}
Here, $E_{01}\simeq 0.2$~meV is the experimental estimate for energy splitting between the logical qubit states~\cite{Shi:2011p233108}.  $t_1$ and $t_2$ are the quantum dot tunneling matrix elements.  For the case that all electrons are in their ground-state, single-particle orbitals (as appropriate for valley-type excited states in Si) $t_2=\sqrt{3/2}t_1$~\cite{Shi:2011p233108}. %
The decoherence model we use is appropriate for charge-state dephasing in a tunnel-coupled double quantum dot~\cite{Barrett:2002p125318}, where $\Gamma=1/T_2^* \simeq 0.2$~GHz is the experimentally measured value for charge qubits in GaAs~\cite{Petersson:2010p246804%
}, and $\gamma \simeq 1$~MHz is the theoretical estimate for $1/T_2^*$, far from the anti-crossings~\cite{Gamble:2012p035302}.

Fig.~\ref{fig:fidelity} shows the results of our dynamical simulation for the worst-case scenario of a $\pi$ rotation around the logical $x$ axis, using an equivalent gate sequence $U=P_2BA(\pi)BP_1$ (see supplemental material).
($Z$ rotations can be achieved with much greater fidelity, since they can be performed
without transforming into state $\ket{E}$.)
Increasing the tunnel coupling improves the fidelity because it increases the speed
of rotation, until
the $A$ and $B$ anti-crossings overlap, at which point the fidelity flattens out.  The point at which this occurs moves to higher frequency as $E_{01}$ increases.

Fig.~\ref{fig:fidelity} also shows analogous fidelity calculations for the exchange
gate that implements $z$ rotations of singlet-triplet qubits~\cite{Petta:2005p2180,Maune:2012p344}, which
are implemented by pulsing to a value of the detuning $\varepsilon$ at which
the exchange coupling $J$ dominates over the inter-dot magnetic field difference $\Delta B$~\cite{Petta:2005p2180}.
There are competing effects in the fidelity when $\Delta B \ll J$ (\emph{i.e.}, when $|\varepsilon|$ is small):
the qubit becomes charge-like, and decoheres more quickly, but
the gate speed also increases.
In Fig.~\ref{fig:fidelity}, the value of $\varepsilon$ is chosen to yield the maximum
value of the fidelity
for every $\Delta B$ and $t_1$ in the simulations.
Figure~\ref{fig:fidelity} shows our results for three physical systems:  GaAs ($\gamma = 1/T_2^*=0.14$~GHz, $\Delta B=3.6$~mT~\cite{Assali:2011p165301}), natural Si ($\gamma = 1.5$~MHz, $\Delta B=26$~$\mu$T~\cite{Assali:2011p165301}), and isotopically purified Si ($\gamma = 0.2$~MHz, $\Delta B=1.2$~$\mu$T~\cite{Assali:2011p165301}).
For a fixed tunnel coupling, increasing $\Delta B$ reduces the fidelity of the exchange gate.
However, better fidelities can be achieved by increasing $\Delta B$ and $t_1$ simultaneously.
Fig.~\ref{fig:fidelity} demonstrates that the fidelity of a pulse-gated hybrid qubit is
comparable to that of a pulse-gated singlet-triplet qubit fabricated using natural silicon.

A different version of the pulse-gated quantum operation can be performed
using a combination of slow ramps and fast pulses
that yield adiabatic passage through the $B$ anticrossing
between $\ket{E}$ and $\ket{1}_L$~\cite{Nakamura:2001}
and Rabi oscillations at the anticrossing $A$ between
$\ket{0}_L$ and $\ket{E}$.
Starting from a detuning that is more negative than $\varepsilon_B$, one first
increases the detuning adiabatically through anticrossing $B$ (which
transforms $\ket{1}_L \rightarrow \ket{E}$ and has no effect on $\ket{0}_L$), 
then pulses
suddenly to anticrossing $A$ (which induces Rabi oscillations between
$\ket{0}_L$ and $\ket{E}$), and finally decreases the detuning adiabatically 
through anticrossing $B$ (which
transforms $\ket{E} \rightarrow \ket{1}_L$ and has no effect on $\ket{0}_L$).
Using a protocol with these adiabatic portions could be very useful if the energy splitting at anticrossing $B$ is significantly larger
than for anticrossing $A$, which is conceivable because of the
large differences of tunnel rates from different orbital states that have been observed
in a silicon quantum dot~\cite{Simmons:2011p156804}.
However, the qubit is much more susceptible to charge noise during
the gating process, because of the markedly different charge distribution
in $\ket{E}$ than in $\ket{0}_L$ and $\ket{1}_L$~\cite{Gamble:2012p035302}, and thus it is likely to be more
difficult to perform high-fidelity gate operations using a partially adiabatic process
than using the sequence of Rabi oscillations described above.

\begin{figure}[t]
\includegraphics[width=3.0in]{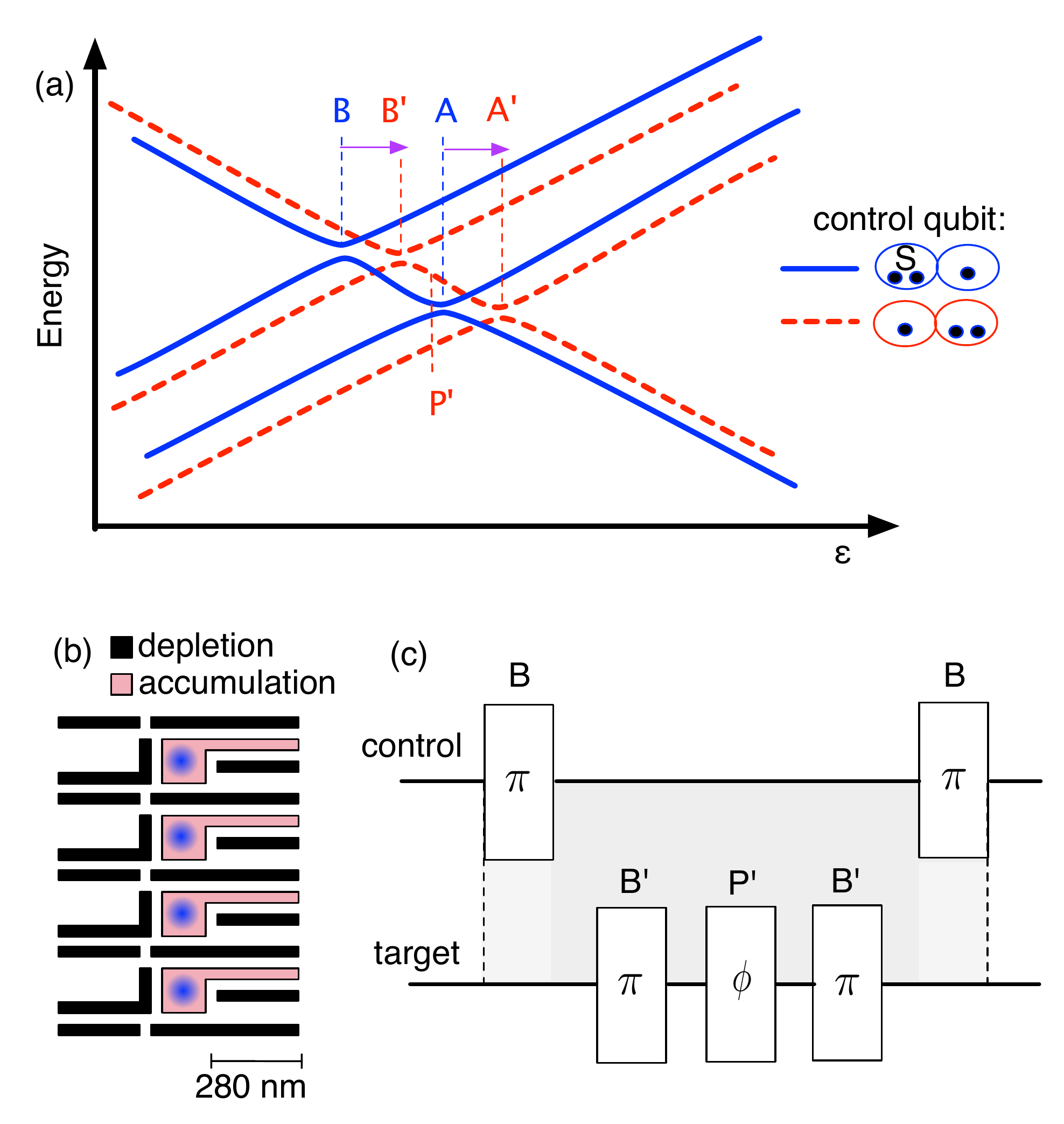}
\caption{\label{fig:two_qubits} 
Illustration of a pulse-gating protocol for a two-qubit gate.
(a) The energy level diagram of target qubit depends on
whether the control qubit is
in state $\ket{0}_L$ or in state $\ket{E}$. 
(The state $\ket{E}$ of the control qubit is obtained from $\ket{1}_L$
by applying the B gate to it.)
The detunings of the avoided crossings A and B are shifted
by an amount $\delta\epsilon$
(denoted by the horizontal arrows on the figure)
when the control qubit is in $\ket{E}$ compared to when the
control qubit is in $\ket{0}_L$.
(b) Realistic device geometry for a top-gated
Si/SiGe heterostructure described in the Supplemental Information.  
2D Thomas-Fermi modeling  \cite{Stopa:1996p13767} of this device
yields $\delta\epsilon\simgt 0.1$ meV, which is ample for operation of a conditional gate.
(c) Gate sequence for performing a conditional rotation of the phase
of the target qubit.  One first applies a B gate to the control qubit, which transforms
$\ket{1}_L\rightarrow\ket{E}$.
One then pulses the target qubit to anticrossing $B^\prime$, then to 
$P^\prime$, and then
again to $B^\prime$, which
changes the phase of the target qubit
only when it starts
in state $\ket{1}_L$ and the control qubit is in state $\ket{E}$.
Finally, one converts the control qubit back to a spin qubit by
applying a $B$ gate to transform $\ket{E}\rightarrow\ket{1}_L$.
The gray shading denotes the conditional nature of the operations
on the target qubit between application of the two B gates to the control qubit.
This gate sequence yields a conditional gate, since it changes the phase 
of the target qubit only when the control qubit starts in state $\ket{1}_L$.
The operations also perform a conditional phase rotation on the control
qubit, which, if desired, can be adjusted to be a multiple of $2\pi$ by
appropriate choice of pulse amplitudes and lengths.
}
\end{figure}

\emph{Two-qubit gates.}---Two-hybrid qubit gates can be implemented by
exploiting
capacitive coupling~\cite{Yamamoto:2003p941,Shulman:2012p202},
as illustrated in Fig.~\ref{fig:two_qubits}.
The charge distribution in state $\ket{E}$ is substantially different than in $\ket{0}_L$,
so there is a substantial Coulomb coupling that causes the location of the anticrossings
$A$ and $B$ 
of the target qubit to depend on
whether the control qubit is in state $\ket{0}_L$ or in state $\ket{E}$,
as shown in Fig.~\ref{fig:two_qubits}.
Therefore, pulsing the target qubit to the detuning of anticrossing A
converts the state $\ket{1}_L$ of the target qubit to $\ket{E}$
when the control qubit is in state $\ket{0}_L$
but not in state $\ket{E}$.
This dependence of the position of the anticrossings of the target qubit on the
state of the control qubit enables the construction of a conditional two-qubit gate, as
illustrated in Fig.~\ref{fig:two_qubits}(c).
One first applies a B gate to the control qubit, which transforms 
$\ket{1}_L\rightarrow \ket{E}$, and then applies a gate sequence that
changes the phase of the target qubit only if the control qubit is in state $\ket{0}_L$.
2D Thomas-Fermi modeling  \cite{Stopa:1996p13767} of the realistic device geometry shown in Fig.~\ref{fig:two_qubits}(b) and described in the supplemental information
yields shifts in the anticrossing
energies of $\simgt 0.1$meV, large enough for fast operations to be feasible.

\emph{Summary and Conclusions.}---In summary, we have presented a method for performing pulse-gating on
a hybrid qubit.
The protocol is more complicated than for a charge qubit
because the qubit states typically cannot be made energetically degenerate.  We overcome this difficulty
by exploiting avoided crossings at two
different detunings between each of the two
qubit states and an intermediate state.
By introducing an additional phase gate at a third detuning point, we have shown that it is possible to implement arbitrary rotations of the logical qubit.
We have derived a closed set of equations for the pulse sequences and performed dynamical simulations of the gates assuming realistic values for the dephasing.
We also showed that two-qubit gates can be implemented by operating the control qubit in the charge regime to electrically enable or disable a rotation on the target qubit.

%
We thank Xuedong Hu, Jon Prance, and Zhan Shi for useful conversations.  This work was supported in part by ARO (W911NF-08-1-0482) and NSF (DMR-0805045, PHY-1104660), and the National Science Foundation Graduate Research Fellowship (DGE-0718123).

\begin{center}{Supplemental Information}\end{center}
\emph{Details of the Derivations of Eqs.~\ref{eq:theta}-\ref{eq:phi2}.}---Here, we outline the calculations of the control parameters $\phi_1$, $\phi_2$, $\theta$ for the single-qubit gate shown
in Fig.~\ref{fig:combined}(c).  

A general rotation of a two-component spinor around the axis $\hat{\mathbf{n}}$ with angle $\phi$ is given by
\begin{equation}
e^{-i\boldsymbol\sigma \cdot \hat{\mathbf{n}} \phi/2} = \cos (\phi/2)
-i\boldsymbol\sigma \cdot \hat{\mathbf{n}} \, \sin  (\phi/2) , \label{eq:rotate}
\end{equation}
where $\boldsymbol\sigma =(\sigma_x,\sigma_y,\sigma_z)$ are Pauli matrices.
The $B$ gate corresponds to an $x$ rotation of angle $\pi$ in the $\{ \ket{1}_L ,\ket{E} \}$ subspace of the full Hilbert space spanned by $\{ \ket{0}_L,\ket{1}_L,\ket{E} \}$.
During the course of the operation, states $\ket{1}_L$ and $\ket{E}$ gain a phase of $e^{i\phi_B}$.
(We define all phases relative to $\ket{0}_L$.)
Thus, 
\begin{equation*}
B=\begin{pmatrix}
1 & 0 & 0 \\
0 & 0 & -i e^{i\phi_B} \\
0 & -i e^{i \phi_B} & 0
\end{pmatrix} . 
\end{equation*}
The $A$ gate corresponds to an $x$ rotation of angle $\theta$ in the $\{ \ket{0}_L ,\ket{E} \}$ subspace.
During this operation, state $\ket{1}_L$ gains a phase of $e^{i\alpha_A}$.
Thus,
\begin{equation*}
A=\begin{pmatrix}
\cos (\theta/2) & 0 & -i \sin (\theta/2) \\
0 & e^{i\alpha_A} & 0 \\
-i \sin (\theta/2) & 0 & \cos (\theta/2)
\end{pmatrix} .
\end{equation*}

The $P_1$ and $P_2$ gates are phase gates.
During the $P_1$ operation, state $\ket{E}$ gains a phase of $e^{i\phi_1}$ relative to $\ket{0}_L$, while state $\ket{1}_L$ gains a phase of $e^{i\alpha_1}$.
Similar considerations apply to $P_2$
Thus,
\begin{equation*}
P_{1,2}=\begin{pmatrix}
1 & 0 & 0 \\
0 & e^{i\alpha_{1,2}} & 0 \\
0 & 0 & e^{i\phi_{1,2}}
\end{pmatrix} .
\end{equation*}

Computing the composite gate $U=BP_2AP_1B$, we obtain the following rotation in the $\{ \ket{0}_L,\ket{1}_L \}$ subspace:
\begin{equation*}
R=\begin{pmatrix}
\cos (\theta/2) & -e^{i(\phi_1+\phi_B)} \sin (\theta/2) \\
-e^{i(\phi_2+\phi_B)} \sin (\theta/2) & -e^{i(\phi_1+\phi_2+2\phi_B)}\cos (\theta/2) 
\end{pmatrix} .
\end{equation*}
We can decompose this into a sum of Pauli matrices using $R=\sum_{j=0,x,y,z} C_j \sigma_j $,
where $\sigma_0=I$ is the identity matrix, and 
$ C_j=\text{Tr}[\sigma_j R]/2 $.
Comparing to Eq.~(\ref{eq:rotate}), we obtain the results shown in Eqs.~(\ref{eq:theta})-(\ref{eq:phi2}), up to an overall phase.

We note that an alternative and equivalent pulse sequence is obtained when $\varepsilon_P<\varepsilon_A$, corresponding to the gate sequence $U=P_2BABP_1$.
For the this gate sequence, we obtain identical results after making the substitutions $\alpha_{0,1}\leftrightarrow\phi_{0,1}$.

\emph{Description of the modeled device.}---The Thomas-Fermi calculations are performed on a realistic quadruple quantum dot gate geometry for an accumulation-mode device.  Fig.~\ref{fig:two_qubits}(b) shows the primary gate pattern, in which metallic gates sit on a 10~nm thick Al$_2$O$_3$ layer on top of a Si/Si$_{0.7}$Ge$_{0.3}$ heterostructure containing a Si quantum well 35~nm beneath its surface.  The Si$_{0.7}$Ge$_{0.3}$ layer is modeled as a dielectric with $\epsilon = 13.19$.  The accumulation gates are positively biased, resulting in electron accumulation as indicated schematically on the diagram.  The depletion gates provide tunability of tunnel barriers between the dots themselves and between each dot and its reservoir.  Two large top gates sit on top of an additional 70~nm of Al$_2$O$_3$.  The first, positively biased, establishes the reservoirs on the left-hand side of Fig.~\ref{fig:two_qubits}(b); the second, negatively biased, prevents any undesired accumulation on the right-hand side of Fig.~\ref{fig:two_qubits}(b).

\bibliography{siliconQCsnc}

\end{document}